\documentclass[pra, reprint, 10pt, british,unsortedaddress]{revtex4-1}
\usepackage[T1]{fontenc}
\usepackage[latin9]{inputenc}
\usepackage{geometry}
\usepackage{color}
\usepackage{units}
\usepackage{amssymb}
\usepackage{graphicx}
\usepackage{esint}

\makeatletter

\providecommand{\tabularnewline}{\\}

\makeatother

\usepackage{babel}
\begin{document}

\title{Maximum information photoelectron metrology}

\author{P. Hockett}

\email{paul.hockett@nrc.ca}

\affiliation{National Research Council of Canada, 100 Sussex Drive, Ottawa,K1M
1R6, Canada}

\author{C. Lux }

\affiliation{Institut für Physik, Universität Kassel, Heinrich-Plett-Str. 40,
34132 Kassel, Germany}

\author{M. Wollenhaupt}

\affiliation{Institut für Physik, Carl von Ossietzky Universität Oldenburg, Carl-von-Ossietzky-Straße
9-11, 26129 Oldenburg, Germany}

\author{T. Baumert}

\affiliation{Institut für Physik, Universität Kassel, Heinrich-Plett-Str. 40,
34132 Kassel, Germany}
\begin{abstract}
Photoelectron interferograms, manifested in photoelectron angular
distributions (PADs), are a high-information, coherent observable.
In order to obtain the maximum information from angle-resolved photoionization
experiments it is desirable to record the full, 3D, photoelectron
momentum distribution. Here we apply tomographic reconstruction techniques
to obtain such 3D distributions from multiphoton ionization of potassium
atoms, and fully analyse the energy and angular content of the 3D
data. The PADs obtained as a function of energy indicate good agreement
with previous 2D data and detailed analysis {[}Hockett et. al., Phys.
Rev. Lett. 112, 223001 (2014){]} over the main spectral features,
but also indicate unexpected symmetry-breaking in certain regions
of momentum space, thus revealing additional continuum interferences
which cannot otherwise be observed. These observations reflect the presence of additional ionization pathways and, most generally, illustrate the power of maximum information measurements of coherent observables for quantum metrology of complex systems.
\end{abstract}
\maketitle

\section{Introduction}

Interferometric measurements are the gold-standard in metrology, since
they offer observables of high precision and information-content,
which can be used to glean detailed understanding of underlying physical
processes. In particular, interference patterns can be used to obtain
the relative phase(s) of contributing waves, quantities which provide
key physical insights in general, and in particular provide a window
into underlying quantum mechanical phenomena. One specific and well-known
example is the wave nature of a free electron, as verified by Young's
double-slit type experiments. In such experiments, a particle described
by a plane-wave impinges on two slits in an otherwise opaque barrier,
resulting in a more complex wave pattern described by two transmitted
spherical wavefronts. These wavefronts interfere with each other,
leading to a characteristic interference pattern. In this case, the
relative phase of the spherical wavefronts depends on the distance
from the slits, and this geometric phase manifests as a spatial dependence
of the observed interferogram.

\begin{figure}
\includegraphics[scale=0.9]{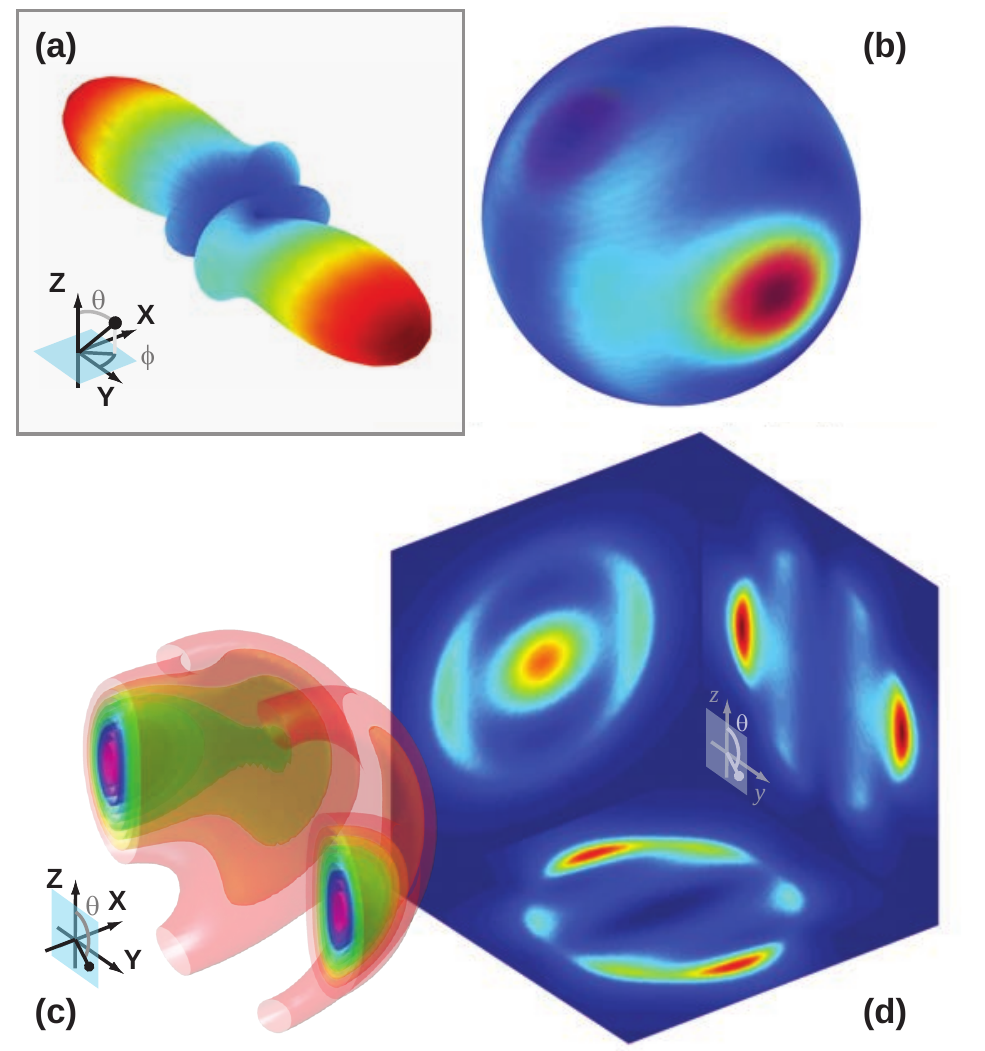}

\caption{Concepts in photoelectron interferometry and imaging. (a) Example
photoelectron interferogram in angular space, $I(\theta,\phi)$, plotted
in polar form. (b) Same angular interferogram as (a), projected onto
the surface of a Newton sphere (velocity iso-sphere), as it would
appear spatially in an ionization experiment for a single photoelectron
energy $k$. (c) Volumetric representation of the angular interferogram
(a) in velocity space (shown only for one hemisphere), assuming a
Gaussian energy envelope. This is essentially a set of nested Newton
spheres (b), now plotted as flux iso-surfaces, and the full distribution
is denoted $I(\theta,\phi,k)$. (d) 2D images of the photoelectron
flux (c), simulating a velocity-map imaging measurement. The interferogram
illustrated is the theoretical result of 3-photon ionization of potassium
with elliptically polarized light (see sects. \ref{sec:Photoelectron-metrology}
for further details), where the laser propagates along the $Z$-axis,
the polarization ellipse lies in the $(X,Y)$ plane with ellipticity
defined by a spectral phase $\phi_{y}=0.5$~rad. (see sect. \ref{sub:Comparison-with-theory}
for details).\label{fig:imaging-concepts}}

\end{figure}

The process of atomic or molecular photoionization is conceptually
similar, and has long been discussed in terms of interfering wavefronts
\cite{Cohen1966,Cooper1968}, but is in general significantly more
complex. An illustrative example is the photoionization of H$_{2}$,
the ``simplest double slit'' \cite{Akoury2007}. In this case, following
single-photon absorption, the photoelectron wavefunction can be considered
as a superposition of two indistinguishable spherical waves, launched
from the two atomic sites  upon photoabsorption. The observable photoelectron
flux, determined by the coherent square of the continuum wavefunction,
exhibits a characteristic angular interference pattern in very close
analogy with Young's double-slit \cite{Cohen1966,Akoury2007}. Another
illustrative example is Young's double slit in the time-domain, which
has been demonstrated by atomic ionization via a coherent two pulse
laser sequence, resulting in an interference pattern in the photoelectron
energy distribution \cite{Wollenhaupt2002,Wollenhaupt2013}. In this
case the temporal evolution of the interferogram responds to the relative
phase of the laser pulses, which can be controlled via their temporal
separation, and is ultimately transferred to the photoelectron wavefunction.

For more complex light-matter systems the continuum wavefunction is
described by a superposition of many constituent \emph{partial-waves}
of differing character, correlated with the various ionization pathways
accessed, and the simple analogy with Young's double-slit fails.
Despite this complexity, the resulting photoelectron flux, measured
spatially, remains, in essence, a self-referencing angular interferogram
of the continuum wavefunction. In this light, measurements of the
energy and angle-resolved photoelectron flux, i.e. the 3D photoelectron
momentum distribution, are particularly powerful, since they provide
a phase-sensitive metrology of the continuum wavefunction and the
scattering event (photoionization) which gave rise to this wave. For
example, photoelectron interferograms have been used to obtain complete
information on the scattering wavefunction for both atoms and molecules
\cite{Duncanson1976,Leahy1991,Becker1998,Reid2003,Gesner2002}; to
investigate electron correlation and entanglement in multiple ionization
\cite{Akoury2007,Schoffler2011}; and in the related context of photoelectron
diffraction \cite{Meckel2008} and photoelectron holography \cite{Huismans2011,Meckel2014},
wherein intense laser fields are used to drive re-scattering of continuum
waves from the photoion and, in the case of holography, additional
interferences between direct (reference) and the re-scattered continuum
waves are observed. Furthermore, in time-dependent cases, the continuum
wavefunction will respond to the underlying dynamics of the ionizing
system, for instance evolving electronic or nuclear configurations
\cite{Assion1996,Arasaki1999,Wollenhaupt2003,Hockett2011,Wang2014},
which may in turn depend on the properties of the laser pulse(s) applied
and even allow for control \cite{Wollenhaupt2009a,Hockett2014}. It
is also of note that the scattering phase accumulated by the partial-waves
is correlated, in the time-domain, with the relative ionization time-delay,
often termed the Wigner delay \cite{Wigner1955,DeCarvalho2002}. In
short, photoelectron interferograms are remarkably versatile and rich,
with many existing and potential applications.

The last decade has seen a surge in ``users'' of photoelectron interferograms,
and as many types of experimental study \cite{Reid2012}. While this
popularity is in part due to the versatility of the measurement, it
is largely due to the proliferation of photoelectron imaging techniques.
In particular \emph{velocity-map imaging} (VMI), since a basic VMI
apparatus offers a robust and simple experimental configuration for
measuring photoelectrons. However, in the majority of cases the use
of standard VMI methodologies means that measurements are restricted
to 2D projections of the full 3D momentum distribution \cite{Parker1997,Parker1997a,Whitaker2003}.
Although 2D projections of any arbitrary 3D distribution may be measured
and interpreted phenomenologically, the resulting loss of dimensionality
means that only cylindrically-symmetric distributions can be quantitatively
analysed \cite{Whitaker2003,Garcia2004}. This further limits the
information extractable from the 2D data and, more fundamentally,
the type of experimental studies possible. While it is trivial to
state that 3D measurements offer a higher information content than
2D measurements, the restrictions inherent to 2D measurements are
highly detrimental to the understanding of the photoelectron interference
pattern, and the concomitant ability to use subtle changes in this
pattern as a probe of the underlying quantum mechanics. This statement
becomes more applicable as the complexity of the light-matter interaction
grows, and the number of interfering pathways increases; for instance
ionization with polarization-shaped laser pulses, where highly-structured,
non-cylindrically symmetric, photoelectron distributions are the norm
\cite{Wollenhaupt2009a,Hockett2014}. 

Figure \ref{fig:imaging-concepts} illustrates some of these concepts
in photoelectron interferometry and imaging, with an example angular
interferogram $I(\theta,\phi)$, representations in velocity space
for (b) narrow and (c) broad energy distributions, and (d) 2D projections
of (c) onto various image planes. The details of these calculations
are discussed further below, but we note here that in this particular
case only the 2D projection in the $(X,Y)$ plane reveals the non-cylindrically
symmetric nature of the distribution, and this key information is
lost in the other projections. While the precise details of the information
loss depend on the initial distribution, and the geometry of the measurement,
this result is applicable to all cases in which the symmetry is broken
in the plane of polarization, e.g. ionization with elliptically polarized
or polarization-shaped laser pulses. Because this plane is orthogonal
to the beam propagation direction it would not be possible to measure
in a standard VMI configuration.

In recent work, refs. \cite{Hockett2014,Hockett2015a}, we explored
a relatively complex ionization process: a net 3-photon ionization
of potassium atoms using moderately intense 800~nm light and a range
of polarizations from linear to circular, and fully polarization-shaped
pulses. We made use of measurements of 2D photoelectron momentum distributions,
combined with the calculation of 2D photoelectron interferograms (including
intra-pulse electronic dynamics, driven by the instantaneous pulse
polarization) and a fitting procedure, in order to determine the full
set of contributing partial-wave magnitudes and phases. Perhaps surprisingly,
this analysis allowed for ``complete'' details of the photoionization
event in terms of the contributing pathways to this particular photoelectron
interferometer, despite the restrictions of the 2D data, but it was
concluded that application to more complex cases would likely require
the additional level of detail available from 3D data; in this work
we explore the capabilities gained from measurement of full 3D photoelectron
distributions created by the same ionization scheme. Measurements
are made using a standard 2D VMI set-up, and combined with tomographic
reconstruction to provide full 3D metrology. With the 3D measurements
we are additionally able to (a) observe non-cylindrically symmetric
distributions directly (sec. \ref{sub:3D-photoelectron-momentum-images})
\footnote{We note for completeness that there are other methods of obtaining
3D distributions, such as direct 3D imaging-type methods which combine
spatial and time-of-flight measurements. Typically such measurements
are more challenging for various reasons, although they are now routine
for a few experimental groups and may offer additional measurement
capabilities such as full photoion-photoelectron vector correlations
depending on the experimental configuration, see for instance refs.
\cite{Continetti2001,Reid2012,Hockett2013} and references therein.
Non-imaging techniques, based on serial point-by-point measurements,
have been in use for decades but are severely limited by the long
collection times required to obtain high resolution 3D datasets, and
may be further limited by the available energy (time-of-flight) resolution.
Some further aspects of 1D, 2D and 3D measurement techniques, including
resolution considerations, are discussed in ref. \cite{Chichinin2009}.
In general we note that 2D imaging combined with tomographic reconstruction
offers a significant reduction in measurement time as compared to
direct 3D imaging techniques, and a simpler experimental set-up.%
}; (b) quantitatively analyse these distributions as a function of
energy (sec. \ref{sub:Angular-data}) and, consequently, (c) compare
these results directly with calculations based on the previously determined
photoionization dynamics (sec. \ref{sub:Comparison-with-theory});
(d) investigate novel symmetry-breaking which was obscured in the
2D measurements, but is clear in the 3D distributions, and directly
reflects additional ionization pathways contributing to the photoelectron
interferogram (sec. \ref{sec:discussion}). All of these aspects serve
to highlight the power of full 3D photoelectron interferograms and
provide a general \emph{maximum information methodology} for analysis
of these measurements, as demonstrated by the new insights obtained
into this complex light-matter interaction.

\section{Photoelectron metrology\label{sec:Photoelectron-metrology}}

In this work ionization of potassium atoms with a single $\sim$30~fs
IR laser pulse was investigated. This provides the specific light-matter
system we use to illustrate the concepts of maximum information photoelectron
metrology. As noted above, interaction of moderately intense ($10^{12}-10^{13}$~Wcm$^{-2}$)
light near 800~nm results in a net 3-photon ionization process. More
specifically, the process can be considered in terms of a strongly-coupled,
bound-bound $4s+h\nu\rightarrow4p_{\pm1}$ transition at the 1-photon
level, followed by a much weaker, 2-photon ionizing transition $4p_{\pm1}+2h\nu\rightarrow|k,l,m\rangle$,
where the continuum states are labelled by photoelectron energy $k$
and (orbital) angular momentum $l$ with projection $m$ on the $Z$-axis.
Because the bound-bound transition is near resonant at 800~nm, and
carries significant oscillator strength, Rabi oscillations are driven;
these intra-pulse electronic population dynamics play a significant
role in the final photoelectron interferogram. Furthermore, since
the polarization of the laser pulse affects both the population dynamics
and the ionization dynamics, the final continuum state populated is
sensitive to the pulse polarization. In effect, the polarization of
the light controls the photoionization interferometer, and finer control
can be gained via the use of polarization shaped pulses, which have
a polarization state that evolves over the pulse envelope. This process
serves to represent a typical, complex, light-matter interaction in
the sense discussed above: many partial-waves contribute to the final
continuum state; electronic dynamics play a significant role; the
interaction is sensitive to controllable experimental parameters,
as well as the inherent physical properties of the ionizing system.
As described above, in this work we are concerned with the additional
insight available from 3D measurements, so the reader is referred
to papers 1 \& 2 for full details of the ionization process, including
the ionization pathways, angular momentum coupling diagrams and the
full theoretical treatment.

In order to develop a maximum information methodology for photoelectron
interferograms we make use of three key elements: (1) sets of 2D VMI
measurements, (2) tomographic reconstruction techniques, (3) detailed
analysis of the radial and angular content of the resulting 3D data,
with a particular focus on the angular photoelectron flux. Each of
these aspects is detailed below, and the data presented in section
\ref{sec:Results-&-analysis}.

\subsection{Experimental set-up\label{sub:Experimental-set-up}}

\textcolor{red}{}

\begin{figure*}
\includegraphics[scale=0.7]{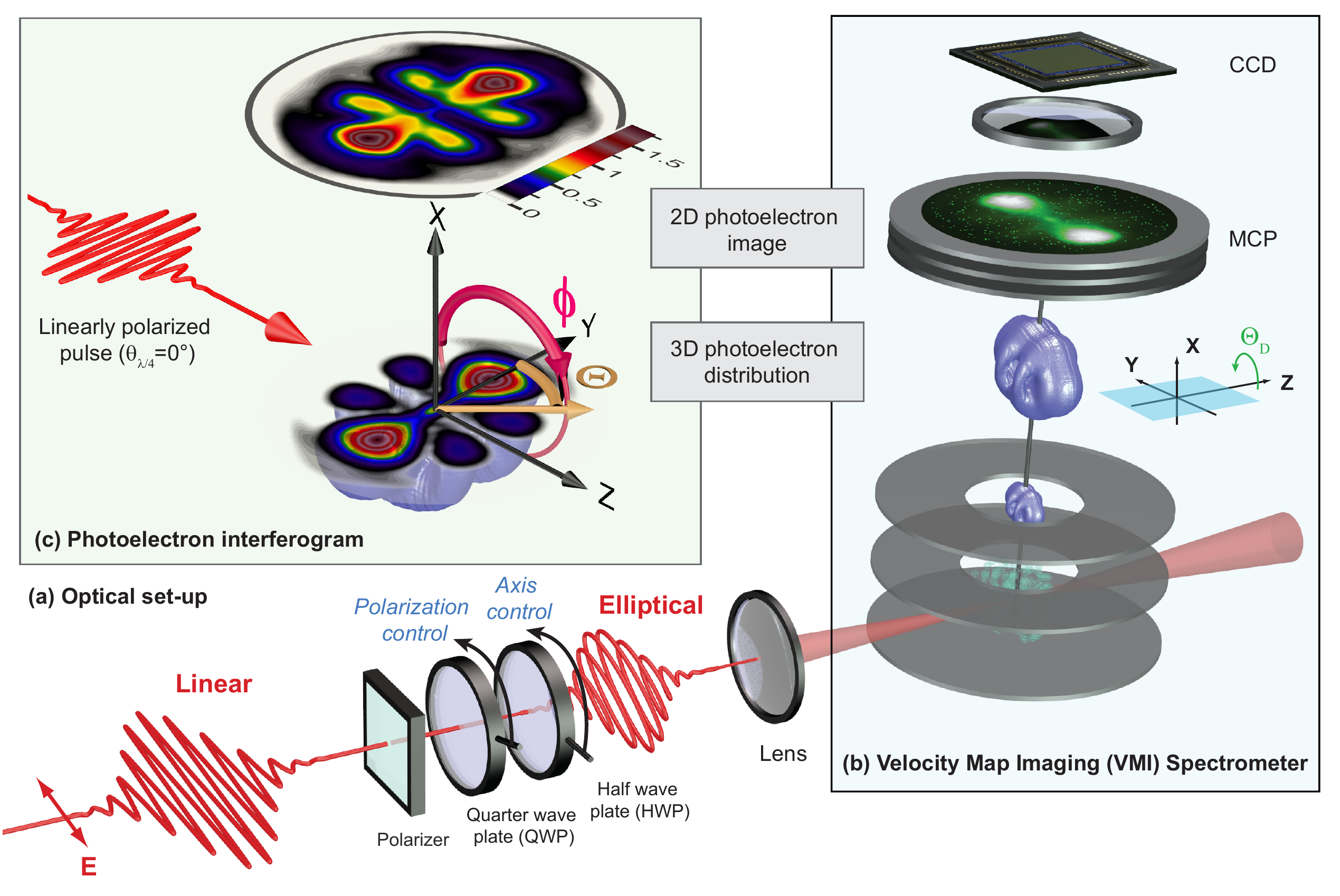}

\caption{Experimental set-up for photoelectron imaging and tomography. (a)
Optical set-up. The polarization state of the incident light is controlled
via a quarter-wave plate (QWP), and rotation relative to the detector
plane via a half-wave plate (HWP). (b) Velocity-map imaging. Potassium
atoms are ionized, and the resulting photoelectron distribution is
projected onto a micro-channel plate (MCP) assembly, allowing 2D images
to be recorded. (c) Photoelectron interferogram measured for ionization
with a linearly polarized pulse, leading to a cylindrically symmetric
distribution. The 2D projection shows an experimental photoelectron
image on the $(Y,Z)$ detector plane, and the 3D distribution reconstructed
from this. This case can be contrasted with the non-cylindrically
symmetric example shown in figure \ref{fig:imaging-concepts}.\label{fig:Experimental}}

\end{figure*}

The experimental set-up is illustrated in figure \ref{fig:Experimental}.
Herein we describe the salient details, and the reader is referred
to refs. \cite{Wollenhaupt2009a} and \cite{Lux2015} for a more detailed
description. In brief, femtosecond laser pulses of $27\,\unit{fs}$
pulse duration (full width at half maximum (FWHM) of the intensity
profile) centered at $795\,\unit{fs}$ with a pulse energy of $800\,\unit{\mu J}$
are provided by an amplified $1\,\unit{kHz}$ Ti:sapphire laser system
(\textit{Femtolasers Femtopower Pro}). The laser beam is focussed
with a lens of 200~mm focal length into potassium vapor supplied
by an alkali metal dispenser source. A mean focal spot size radius
of about $22\,\unit{\mu m}$ ($1/e^{2}$ of intensity profile) was
obtained and measured with a beam profiling CCD camera. At a pulse
energy of $7.8\,\unit{\mu J}$ this leads to a peak intensity of about
$4\cdot10^{13}$~Wcm$^{-2}$ assuming Gaussian profiles in time and
space.

We use a home-built velocity map imaging (VMI) spectrometer \cite{Parker1997}
to record 2D photoelectron images, illustrated in figure \ref{fig:Experimental}.
The imaging assembly consists of a chevron micro-channel plate (MCP)
detector with a phosphor screen deposited on a fiber optic (\textit{SI-Instruments
GmbH model S3075-10-I60-PS-FM}). A $10\,\unit{bit}$ CCD-camera with
1.4 million pixels (\textit{Lumenera Corporation model Lw165m}) is
used to image the signals on the phosphor screen. The energy resolution
of the VMI spectrometer in the present measurements is better than
$80\,\unit{meV}$ (FWHM) at an energy of about $0.5\,\unit{eV}$.

The polarization of the initial laser pulses is linear with the polarization
axis perpendicular to the spectrometer axis and coplanar to the detector
surface, as shown in figure \ref{fig:Experimental}. We use a dichroitic
VIS-IR-polarizer (\textit{CODIXX}) to clean up the linear polarization.
An achromatic quarter-wave plate (QWP) (\textit{B. Halle Nachfl.})
is placed after this polarizer to generate different polarization
states with an adjustable amount of circularity. The polarization
states are therefore defined by the QWP rotation angle $\theta_{\lambda/4}$,
where $\theta_{\lambda/4}=0^{\circ}$ corresponds to linearly polarized
light, $\theta_{\lambda/4}=45^{\circ}$ to circularly polarized light,
and all values in between to elliptically polarized pulses. The ellipticities
of the laser pulses are defined as the ratio of the minor to major
axes of the polarization ellipse, hence $\varepsilon=0$ for linearly
polarized light and $\varepsilon=1$ for pure circularly polarized
light. Herein we present data for three polarization states defined
by $\theta_{\lambda/4}=0^{\circ}$ (linear polarization), $15^{\circ}$
and $30{}^{\circ}$ (elliptically polarized states). The corresponding
Stokes vectors, denoted $S(\theta_{\lambda/4})$, and ellipticities
$\varepsilon(\theta_{\lambda/4})$, are given in table \ref{tab:Polarization-states},
where the Stokes parameters were measured experimentally using the
method of ref. \cite{Schaefer2007} (further details can be found
in ref. \cite{Lux2015}). In this work only the ellipticities are
of fundamental importance, and hereafter the three polarization states
are denoted $\varepsilon_{1}$, $\varepsilon_{2}$ and $\varepsilon_{3}$
respectively. Although the Stokes parameters indicate a slight rotation
of the polarization ellipse as a function of $\theta_{\lambda/4}$
(relative to the lab. frame $Y$-axis defined in figure \ref{fig:Experimental-data}),
this rotation does not affect the photoelectron distributions beyond
a trivial frame rotation, and was removed during data processing (although
is present in the raw data shown in figure \ref{fig:Experimental-data}).

\begin{table}
{\footnotesize{}}%
\begin{tabular}{c|c|c|c}
{\footnotesize{Label}} & {\footnotesize{QWP}} & \multicolumn{1}{c}{Stokes} & {\footnotesize{Ellipticity }}\tabularnewline
 & {\footnotesize{$\theta_{\lambda/4}$}} & {\footnotesize{$S$}} & {\footnotesize{ $\varepsilon$}}\tabularnewline
\hline 
\hline 
{\footnotesize{$\varepsilon_{1}$}} & {\footnotesize{0$^{\circ}$}} & {\footnotesize{(1,~1.00,~0.04,~0.00)}} & {\footnotesize{0}}\tabularnewline
\hline 
{\footnotesize{$\varepsilon_{2}$}} & {\footnotesize{15$^{\circ}$}} & {\footnotesize{(1,~0.76,~0.43,~0.49)}} & {\footnotesize{$\approx$0.27}}\tabularnewline
\hline 
{\footnotesize{$\varepsilon_{3}$}} & {\footnotesize{30$^{\circ}$}} & {\footnotesize{(1,~0.28,~0.39,~0.88)}} & {\footnotesize{$\approx$0.58}}\tabularnewline
\end{tabular}{\footnotesize \par}

\caption{Polarization states used in this work, defined by the QWP rotation
angle $\theta_{\lambda/4}$, and corresponding Stokes parameters and
ellipticities.\label{tab:Polarization-states}}
\end{table}

To record different 2D projections of the photoelectron distribution,
an achromatic half-wave plate (HWP) (\textit{B. Halle Nachfl.}) is
utilized after the QWP. This HWP provides control over the polarization
axes of the incoming light, hence the rotation of the 3D photoelectron
distribution relative to the detector plane. This rotation angle is
defined as $\Theta_{D}$, where $\Theta_{D}=0^{\circ}$ for the case
shown in figure \ref{fig:Experimental}, with the $(Y,Z)$ plane parallel
to the detector plane, and $\Theta_{D}=90^{\circ}$ for the case where
the $(X,Z)$ plane is parallel to the detector (see also figure \ref{fig:Experimental-data}).
Photoelectron images are measured for a range of $\Theta_{D}$, and
this series of projections is used to reconstruct the initial 3D distribution
for each polarization state $\varepsilon_{n}$, as detailed in sec.
\ref{sub:Tomographic-reconstruction}. For each $\Theta_{D}$ images
are integrated for approx. $30000$ laser pulses; under the conditions
described above approx. $33$ electrons are measured per pulse, resulting
in a total electron count of approx. $9\times10^{5}$ per 2D projection.
Overall, each tomographic measurement includes approx. $2.4\times10^{6}$
laser pulses and approx. $8\times10^{7}$ electrons, and takes around
50 minutes. Due to the large number of electron counts in each measurement,
statistical (Poissonian) uncertainties are negligible in this data,
and are consequently not shown on the plots presented herein.

\subsection{Tomographic reconstruction\label{sub:Tomographic-reconstruction}}

\begin{figure}
\includegraphics[scale=1.15]{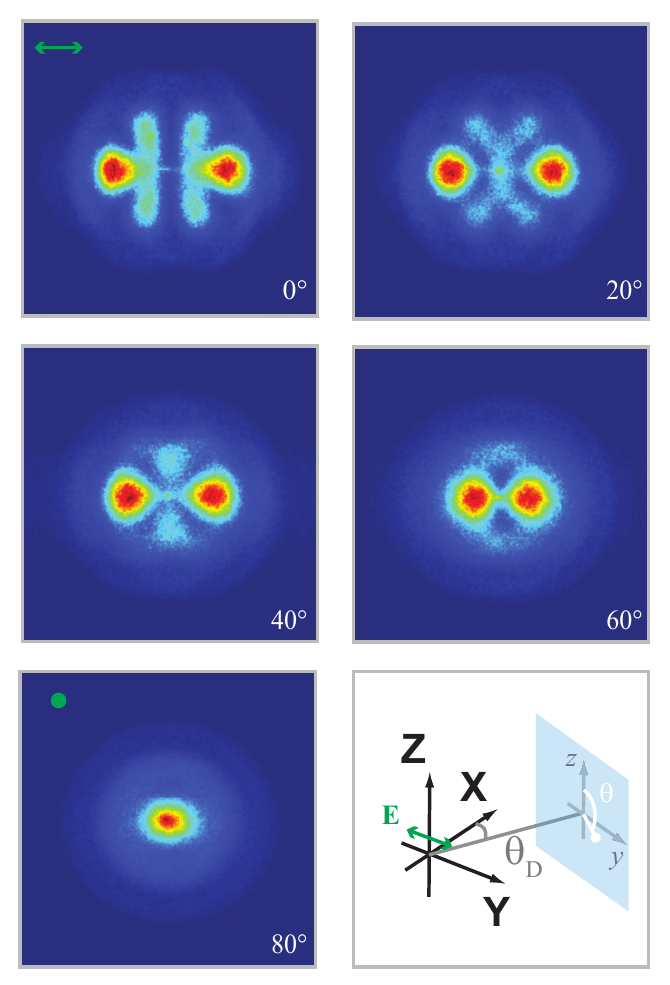}

\caption{Experimental data for photoelectron tomography. Panels show examples
of raw photoelectron images for different projection angles $\Theta_{D}$
onto the 2D imaging detector, following the coordinate system illustrated
in the bottom right panel. The raw images were obtained with linearly
polarized light ($\varepsilon_{1}$), and the frame rotation from
top to bottom corresponds to a rotation of the detection plane $(y,z)$
from $(Y,Z)$ ($\Theta_{D}=0^{\circ}$) to $(X,Z)$ ($\Theta_{D}=90^{\circ}$),
with similar behaviour to the calculated case shown in fig. \ref{fig:imaging-concepts}(d)
for an elliptically polarized ionizing pulse.\label{fig:Experimental-data}}

\end{figure}

For each polarization state, photoelectron images were recorded for
a set of projection angles $\Theta_{D}$, as described above. As illustrated
in figure \ref{fig:imaging-concepts}(c) \& (d), each image is a 2D
projection $(y,z)$ of the original 3D velocity space, $(V_{X},V_{Y},V_{Z})$,
where the lower-case coordinates $(y,z)$ are defined in the detector
plane and the upper-case coordinates $(X,Y,Z)$ in the ionization
frame, where the $Z$-axis is chosen to coincide with the laser propagation
axis. Raw experimental images are shown in figure \ref{fig:Experimental-data}
for a few values of $\Theta_{D}$. In this case, as distinct from
the example shown in figure \ref{fig:imaging-concepts}, the ionizing
light was linear, and the full 3D distribution is cylindrically-symmetric.
Here the symmetry is clear for the $\Theta_{D}=0^{\circ}$ case, which
shows a distribution with two intense poles aligned with the $Y$-axis,
and band structure of lower intensity. As this distribution is rotated
the poles, in 2D projection, appear to come closer to one another,
and ultimately align for $\Theta_{D}=90^{\circ}$. The band structure
initially becomes more complex in projection, and at $\Theta_{D}=90^{\circ}$
appears as a weak, radially dependent feature dropping off from the
central spot. (The reconstructed 3D distribution from this data is
shown in figure \ref{fig:Experimental-3D-photoelectron}.) 

In the tomographic reconstruction procedure the images were stacked
to form a data cube of dimensions $(V_{x},V_{y},\Theta_{D})$, and
an inverse Radon transform performed on each $(V_{y},\Theta_{D})$
plane in the image stack to recreate the original $(V_{X},V_{Y},V_{Z})$
space. This procedure is equivalent to those detailed in refs. \cite{Wollenhaupt2009,Smeenk2009,Hockett2010},
although the numerical details are slightly different. In this case,
the reconstruction was performed in Matlab (R2010a), using the built-in
iRadon function. The input $(V_{x},V_{y})$ images were cropped and
down-sampled by a factor of two before reconstruction, yielding raw
velocity space images of 251x251 pixels, and the inverse Radon transform
included a Ram-Lak frequency filter with Hann windowing to remove
high-frequency noise. Sets of images for $\Theta_{D}=0^{\circ}$ to
90$^{\circ}$, in 2$^{\circ}$ steps, were used, resulting in image
sets of 45 projections for each polarization state. Combined with
the down-sampling, this resulted in a data cube $(V_{x},V_{y},\Theta_{D})$
of dimension 251x251x45, and a reconstructed velocity space volume
of dimension 251x251x251 voxels.

\subsection{Information content of 3D data\label{sub:Information-content}}

In the following we determine and discuss the details of the radial
and angular components obtained from the 3D momentum data. Most generally
the distributions, in spherical polar coordinates, can be described
by:

\begin{equation}
I(\theta,\phi,k)=\sum_{L,M}\beta_{L,M}(k)Y_{L,M}(\theta,\,\phi)\label{eq:IBlm}
\end{equation}

Here the $Y_{L,M}(\theta,\,\phi)$ are spherical harmonic functions
and the $\beta_{L,M}(k)$ are the anisotropy parameters, explicitly
given as a function of energy. The experimental volumetric data can
be expressed in terms of this characteristic expansion by defining
a coordinate origin, converting the Cartesian volume $(X,Y,Z)$ to
a spherical-polar coordinate system $(\theta,\phi,R)$, then extracting
radial slices and, finally, determining the $\beta_{L,M}(k)$ by fitting
the radial slices with eqn. \ref{eq:IBlm}. Note that since the raw
radial spectrum recorded via a VMI experiment is linear in velocity
space, it is non-linear in energy space (because $k\propto v^{2}\propto R^{2}$).
For simplicity of data analysis we therefore work primarily in this
linear space, defined in practice by CCD pixels and labelled by the
arbitrary radial coordinate $R$, but use the notation $\beta_{L,M}(k)$
in all cases. The $\beta_{L,M}(k)$ thus determined constitute the
full information content of the measurement, and fully characterize
distributions such as the one shown in figure \ref{fig:imaging-concepts}(c). 

The radial spectrum, summed over all angles, corresponds to the photoelectron
velocity (or energy) spectrum. This component is given by the $\beta_{0,0}(k)$
parameters, or equivalently can be obtained by direct angular integration
of the data:

\begin{equation}
I(k)=\int\int I(\theta,\phi,k)\sin(\theta)d\theta d\phi
\end{equation}

The angular photoelectron interferograms $I(\theta,\phi)$ at any
given $k$, hence the set of $\beta_{L,M}(k)$ at any given $k$,
depend on the composition of the continuum wavefunction, defined by
the partial-waves $|l,\, m\rangle$. Although the nature of the interferences
and coupling is complicated (see papers 1 \& 2), in general the limits
on $L$ and $M$ in this expansion depend directly on the continuum
states populated, which depend in turn on both the characteristics
of the ionizing radiation and the intrinsic properties of the ionizing
system. These properties effectively determine the ionization pathways
accessible, via the symmetry of the problem and coupling to the partial-waves
\cite{Yang1948,Reid2003,Reid2012}; for example, $L\leq2l_{max}$
and, in the case of linearly polarized light, only $M=0$ terms are
allowed. The resulting interference pattern, at a single energy, is
often termed the \emph{photoelectron angular distribution} (PAD),
as distinct from the radial component which reflects the photoelectron
energy spectrum. It is important to note that the $\beta_{L,M}(k)$
parameters cannot be determined from 2D images in general, due to
the loss of information which occurs with projection of the full distribution
on a 2D plane. 

The exception to this is cylindrically symmetric ($\phi$-invariant)
distributions, for which Abel-type inversion techniques can be employed
\cite{Whitaker2003}. Since cylindrical symmetry is only maintained
for linear or pure circularly polarized light ($\varepsilon=$0 or
$\varepsilon=$1), this stipulation corresponds in practice to a restriction
on the experiments possible and, ultimately, on the partial-wave interferences
which can be observed \cite{Leahy1992,Reid2003}. This latter consideration
provides a fundamental limit to the information content which can
be quantitatively obtained from a 2D measurement.

\textcolor{red}{}

\section{Results \& analysis\label{sec:Results-&-analysis}}

\subsection{3D photoelectron momentum images\label{sub:3D-photoelectron-momentum-images}}

\begin{figure*}
\includegraphics{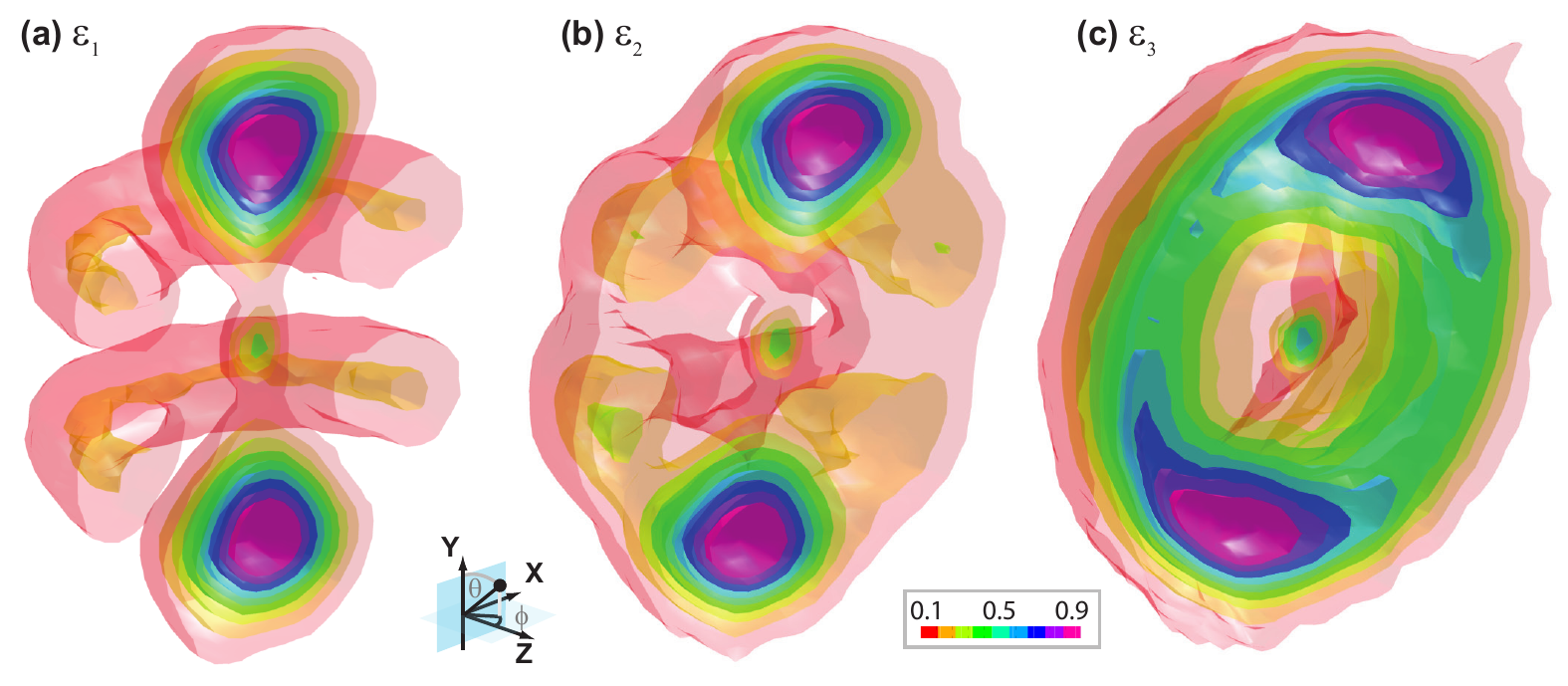}

\caption{Experimental 3D photoelectron images for three polarization states.
(a) Linearly polarized light ($\varepsilon_{1}$); (b) \& (c) elliptically
polarized light $\varepsilon_{2}$ and $\varepsilon_{3}$ respectively
(see table \ref{tab:Polarization-states}). To reveal the details
in the plane of polarization, the distributions are sliced in the
$(X,Y)$ plane, plotted for only one hemisphere and the coordinate
frame rotated as indicated in the figure; the isosurfaces are plotted
for 10-90\% signal levels. \label{fig:Experimental-3D-photoelectron}}

\end{figure*}

The images obtained experimentally are shown in figure \ref{fig:Experimental-3D-photoelectron}.
In the figure we show 3D iso-surface renderings of the tomographically
reconstructed distributions, equivalent to the computational result
shown in figure \ref{fig:imaging-concepts}(c). The distributions
are sliced in the $(X,Y)$ plane to reveal details of the radial (velocity)
spectrum, and highlight the angular structure of the signal in the
polarization plane. The radial component of the images shows two features:
a central spot, and a main radial feature which appears to have a
Gaussian-like envelope. The angular component shows little clear structure
for the central spot, and a complex multi-lobed structure over the
main radial feature.

Broadly, the results over the main feature show the expected behaviour
as the polarization state of the light is changed, evolving from a
distribution with primarily $L=3$ structure (or $f$-like structure
in the language of atomic orbitals %
\footnote{For clarity it is important to note here that while the observed angular
structure may be conveniently labelled with such terms - $s$-like,
$p$-like etc. - the terminology is used in this context only to apply
to the observable angular interferogram, and does not imply any knowledge
about the contributing continuum wavefunction.%
}) in fig. \ref{fig:Experimental-3D-photoelectron}(a), to a more ring-like
structure in fig. \ref{fig:Experimental-3D-photoelectron}(c). This
general evolution with the laser polarization matches that seen in
the 2D photoelectron images recorded at lower laser intensities, and
predicted theoretically (see papers 1 \& 2). There is little apparent
change in the radial distribution with polarization, also as expected,
although more careful analysis (see below) indicates this is not entirely
true. The central spot was not observed in the previous 2D images,
although such a feature is quite typical of VMI measurements and is
usually assumed to indicate photoelectrons generated via many possible
pathways to low-eKE continuum states (e.g. field ionization of high-lying
excited states). Such pathways may be laser intensity and VMI parameter
dependent, and have been exploited in ``photoionization microscopy''
studies \cite{Nicole2000,Lepine2004a}, and more recently investigated
in the context of strong-field atomic and molecular ionization \cite{Wolter2014,Dimitrovski2014}.
In the former case, near threshold ionization is analysed in a joint
atom-electric field potential, leading to the formation of complex
quasi-bound states; in the latter case, low and zero-energy photoelectrons
are associated with electron tunnelling, followed by scattering and
Coulomb focussing or recapture into high-lying Rydberg states, with
subsequent field ionization of these states. Effectively, the same
processes operate in both regimes, but the precise details vary with
the strength of the laser field and the applied static fields. We
do not consider these low-energy contributions further in this work.

In detail, the full 3D data begins to reveal additional information
which can be inferred, but not observed directly, in 2D projections.
As noted above, this is particularly true for any structure in the
$(X,Y)$ plane - the plane of polarization of the laser pulse - which
cannot be observed in a 2D image in standard VMI configurations (see
figures \ref{fig:imaging-concepts} and \ref{fig:Experimental-data}).
For the linearly polarized light the distribution is cylindrically
symmetric so, as described above, there is nominally no loss of information
in the 2D projections. This case does, however, serve as a good test
of the tomographic reconstruction and an intuitive example, since
the distribution is relatively simple and can be readily checked by
eye against the 2D images. It is clear how the $f$-like structure,
with intense polar lobes and two radial bands, can form the various
projections shown in figure \ref{fig:Experimental-data}.

For the elliptically polarized cases the distributions are more interesting;
of particular note is the rotation in the $(X,Y)$ plane, which appears
in the 2D images as a smearing of the features in the equatorial plane.
In this case the rotation is somewhat trivial in origin, and simply
due to the optical set-up (as detailed in section \ref{sub:Experimental-set-up})
which results in a slight reference frame rotation for the elliptically
polarized cases relative to the linear case. In the 3D data this effect
can be clearly observed, understood, and removed in data analysis.
However, it is clear that in a 2D measurement this kind of overlap
or blurring will be present in any case where features overlap in
projection, regardless of the experimental or physical origin of this
overlap. This general effect was illustrated in detail above for the
linear case (figure \ref{fig:Experimental-data}), where the effects
of frame rotations on 2D projections are observed as the approach
and overlap of the polar features as $\Theta_{D}$ is increased. Obviously
the magnitude of this effect, and the loss of detailed structure in
the projections, will depend on the complexity and, especially, the
width of the features. Similarly, the loss of structure away from
the $(X,Y)$ plane can be inferred from a 2D projection as a narrowing
of the observed distribution, but in the 3D case the persistence of
the lobe pattern in plane can be observed. In all cases, the potential
for the conflation of (trivial) experimental with physically interesting
effects is high, and detrimental to detailed analysis and fundamental
insight. 

Although visually arresting and phenomenologically useful, direct
consideration of the iso-surface plots provides only a cursory overview
of the data, and does not clearly reveal minor contributions to the
photoelectron interferograms. In the following sections, a more careful
quantitative analysis of this data is reported in order to provide
more detail, and ultimately provide physical insight into these 3D
interferograms.%
\footnote{A similar analysis of 3D photoelectron interferograms from a time-resolved
case can be found in ref. \cite{Hockett2013}, along with a more detailed
discussion of the dimensionality of the observables.%
}

\subsection{Radial spectra}

\begin{figure}
\includegraphics[bb=0bp 0bp 248bp 216bp]{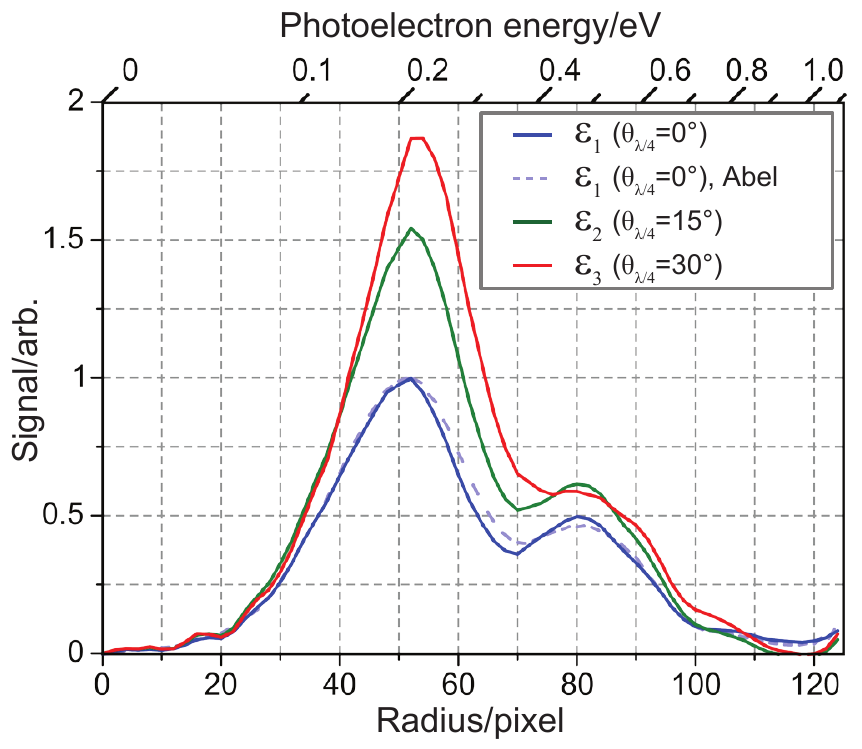}

\caption{Radial spectra obtained from the 3D data shown in figure \ref{fig:Experimental-3D-photoelectron}.
For the linear case ($\varepsilon_{1}$) the spectrum obtained via
the Abel-transform based pBasex algorithm is also shown. The spectra
are plotted in the native velocity space, and the upper abcissca indicates
the (non-linear) calibration to photoelectron kinetic energy. \label{fig:Radial-spectra}}

\end{figure}

The radial spectra extracted from the volumetric data are shown in
figure \ref{fig:Radial-spectra}. As expected from the full 3D plots
of fig. \ref{fig:Experimental-3D-photoelectron}, there is little
difference in the overall structure of the distributions with polarization
state. However, there is a clear double-peak, with a splitting of
$\sim$270~meV. The ratio of the peaks forming this doublet varies
as a function of polarization, with a slight dominance of the lower-energy
feature observed for $\theta_{\lambda/2}=0^{\circ}$ which evolves
smoothly with ellipticity to a more significant dominance at $\theta_{\lambda/2}=30^{\circ}$.
This trend is not obvious from fig. \ref{fig:Experimental-3D-photoelectron},
nor was it observed in previous 2D images. 

The double-peaked structure of the spectrum is due to the Autler-Townes
(AT) effect, and has been studied in detail in previous work \cite{Wollenhaupt2003b,Wollenhaupt2005,Wollenhaupt2009a}.
The dependence of this structure on polarization has not been systematically
explored before; however, since the splitting depends on the time-dependent
Rabi frequency, given by the dipole operator times the (complex) laser
electric field, one might expect it to respond to pulse polarization
due to the dependence of the AT features on the details and ratios
of the various ionization pathways, which are sensitive to the pulse
polarization (see ref. \cite{Hockett2014}). As discussed further
below, there is some evidence for interferences between ionization
paths of different photon-order in the PADs, and this may also contribute
to the differences observed in the radial distributions.

As a further cross-check of the tomographic procedure the radial spectrum
obtained for the cylindrically symmetric $\varepsilon_{1}$ case can
be compared to the spectrum determined using standard Abel-based methods.
Here an adapted version of the pBasex algorithm \cite{Garcia2004,Lux2015}
was used, which makes use of a forward Abel transform of a set of
polar basis functions combined with a fitting routine in order to
determine the original, $\phi$-invariant, photoelectron distribution
$I(\theta,k)$ from a single image (with $\Theta_{D}=0^{\circ}$).
It is clear that the methods are in good agreement.

\subsection{Angular distributions and anisotropy parameters\label{sub:Angular-data}}

\begin{figure}
\includegraphics{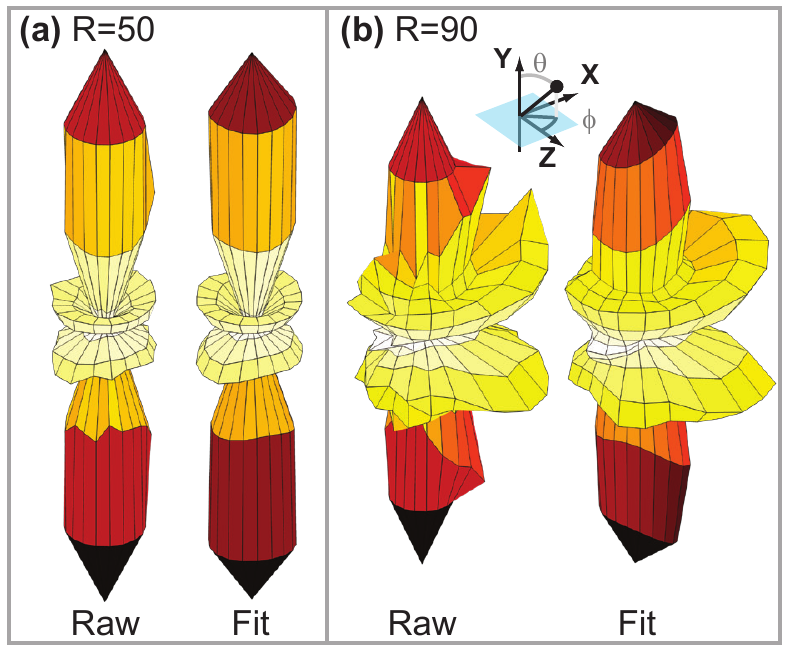}

\caption{Raw \& fitted photoelectron angular interferograms, $I(\theta,\,\phi,\, k)$,
for selected $k$ following ionization with linearly polarized light
($\varepsilon_{1}$). (a) PAD at the peak of the spectrum ($R=50$),
displaying cylindrical symmetry; (b) PAD at the high-energy wing ($R=90$),
displaying symmetry breaking in the plane of polarization. In both
panels the raw data is on the left and the fit on the right, with
the fit results displayed at the same angular binning as used for
the raw data. Further examples of fitted PADs are shown in figure
\ref{fig:PADs_3x5}, and the complete set of extracted $\beta_{L,M}(k)$
in figure \ref{fig:BLM_surfs}. \label{fig:PADs}}

\end{figure}

\begin{figure*}
\includegraphics{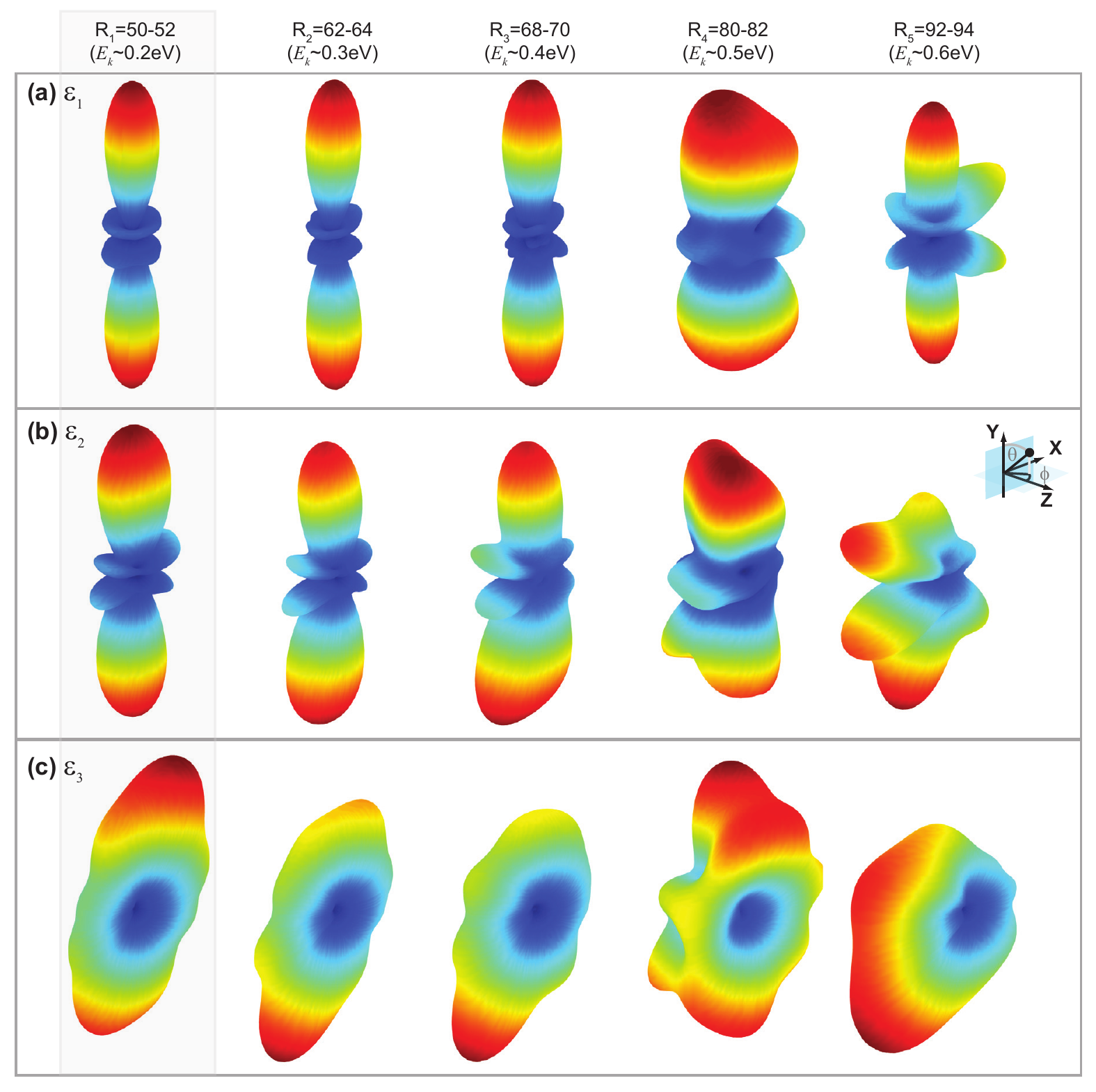}

\caption{Photoelectron angular interferograms. PADs are shown for a range of
photoelectron energies for (a) linearly polarized light, $\varepsilon_{1}$;
(b) \& (c) elliptically polarized light, $\varepsilon_{2}$ and $\varepsilon_{3}$
respectively. Radial windows are labelled 1~-~5, corresponding to
the labels in fig. \ref{fig:BLM_surfs}, and approximate photoelectron
kinetic energies $E_{k}$ are also given. In all cases the PADs are
generated from the $\beta_{L,M}(k)$ obtained from the experimental
data, i.e. correspond to the fit results of figure \ref{fig:PADs}
but at higher-resolution. This is shown explicitly for panel (a),
where the first and last distributions correspond to the distributions
of figure \ref{fig:PADs}. The corresponding $\beta_{L,M}(k)$ are
shown in figure \ref{fig:BLM_surfs}.\label{fig:PADs_3x5}}
\end{figure*}

\begin{figure}
\includegraphics[scale=1.1]{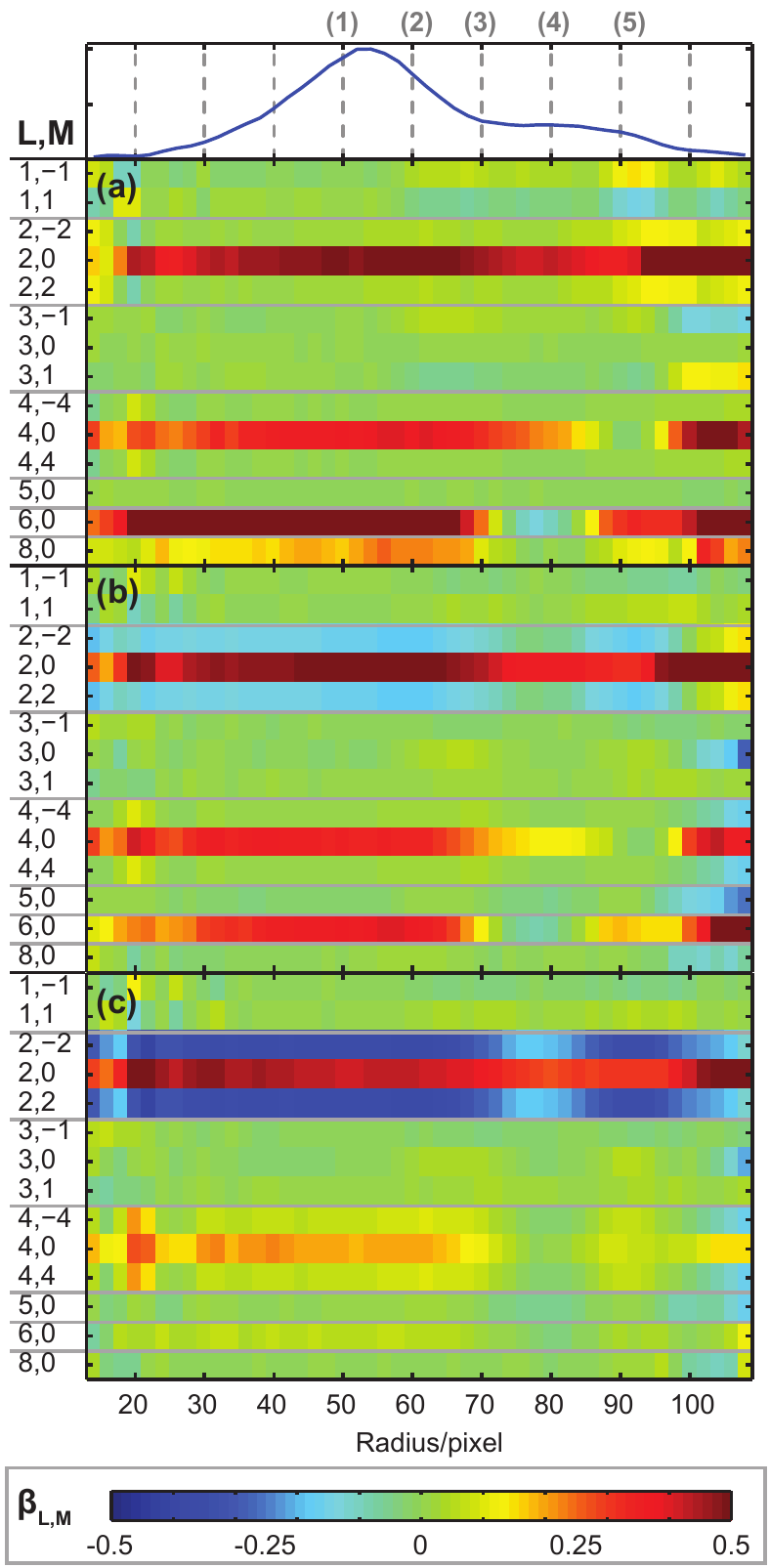}

\caption{Full $\beta_{L,M}(k)$ parameters for (a) linearly polarized light
(a) linearly polarized light $\varepsilon_{1}$; (b) \& (c) elliptically
polarized light $\varepsilon_{2}$ and $\varepsilon_{3}$ respectively.
For reference, the velocity spectrum for the linear case is shown
in the top panel, and the radial slices labelled correspond to the
PADs plotted in figure \ref{fig:PADs_3x5}. All $L,M$ values for
which $|\beta_{L,M}(k)|>0.2$ for any dataset (i.e. all $\varepsilon_{n}$)
are plotted. To best visualize and compare the large parameter space,
the colour bar is fixed over all plots from $-0.5$ to $+0.5$, although
this does result in a few regions over the main feature where the
values are out of range. \label{fig:BLM_surfs}}
\end{figure}

Photoelectron angular distributions (PADs), $I(\theta,\phi,k)$, were
extracted from the 3D data as defined in section \ref{sub:Information-content}.
Figure \ref{fig:PADs} shows some examples of the raw data and the
fitted interferograms, plotted in spherical polar space as a function
of energy. A radial step size of $\Delta r=2$ was used in this procedure,
and the fitting can be considered as both a means to determine the
$\beta_{L,M}(k)$ parameters and a data smoothing procedure. From
these two examples it is apparent that the spherical harmonic expansion
describes the experimental data very well. Furthermore, it is immediately
clear that the PADs have quite complex angular and energy structure,
where only the former aspect was expected from the previously obtained
2D images (ref. \cite{Hockett2014}) and visual inspection of the
3D data (sect. \ref{sub:3D-photoelectron-momentum-images}). Figure
\ref{fig:PADs}(a) corresponds to the main spectral peak. Here the
structure matches that observed directly in the volumetric plots of
figure \ref{fig:Experimental-3D-photoelectron}, and indicates the
expected cylindrical-symmetry. However, figure \ref{fig:PADs}(b),
which corresponds to the high velocity wing of the spectrum, shows
that the form of the PADs changes and, most interestingly, symmetry
breaking is observed, with asymmetries appearing in the plane of polarization.

A more detailed view of this behaviour is presented in figures \ref{fig:PADs_3x5}
and \ref{fig:BLM_surfs}. The former shows PADs for all polarization
states and a range of $k$, while the latter shows the $\beta_{L,M}(k)$
which provided the full information content of the data. Although
not as evocative as the spherical polar representations shown in figure
\ref{fig:PADs_3x5}, this reduced dimensionality representation allows
for a more detailed overall view of the energy-resolved interferograms.
To reduce the complexity of the presentation slightly, the plots here
show only the major parameters, defined in this case by $|\beta_{L,M}(k)|>0.2$
for any polarization state $\varepsilon_{n}$. To maintain consistency
the same set of parameters is shown for all polarization states, and
the colour mapping is similarly maintained for all datasets. Note
that the polar coordinate space used is defined such that the polar
axis is parallel to the $Y$-axis (same convention as figures \ref{fig:Experimental-data},
\ref{fig:PADs} and \ref{fig:PADs_3x5}). This allows for the linear
case to take its simplest form, with $M=0$ terms only, and matches
the frame definition used for the calculations of refs. \cite{Hockett2014,Hockett2015a}.

In both figures \ref{fig:PADs_3x5} and \ref{fig:BLM_surfs} the evolution
of the interferograms with $k$ is clear: over the main spectral feature
($30\lesssim R\lesssim70$) the $\beta_{L,M}(k)$ are almost invariant,
but away from this feature there are significant - but smooth - changes.
In particular, the wings of the peaks ($R\lesssim30$, $R\gtrsim95$)
show significant evolution of the $\beta_{L,M}(k)$, as does the region
of overlap of the major and minor spectral features ($75\lesssim R\lesssim85$).
In general the results show fairly smooth evolution of the $\beta_{L,M}(k)$
with energy, with faster evolution as a function of energy in the
regions of peak overlap. The symmetry breaking noted above corresponds
to the highest energy regions, in $\beta_{L,M}(k)$ terms these are
regions where significant odd-$L$ and odd-$M$ values are present.
This symmetry breaking is particularly clear for cases $\varepsilon_{1}$
and $\varepsilon_{2}$, which both have lobes along the $X$-axis
in the positive or negative directions only, corresponding to significant
$L=1$ and $L=3$ contributions in the $\beta_{L,M}(k)$ spectra.
Additionally, figure \ref{fig:BLM_surfs}(a) also shows significant
$L=8$ terms are present. The interpretation of these observations
is discussed in section \ref{sec:discussion}.

Overall, the results clearly show the benefits of a maximum information
metrology approach. In particular the clear symmetry-breaking within
the plane of polarization is a signature of the presence of additional
interfering channels in the photoionization interferometer. These
signatures are not observable in the current or previous 2D images,
nor allowed by the theoretical treatment, and indicate additional
complexities in the light-matter interaction beyond the net 3-photon
ionization framework previously established (section \ref{sec:Photoelectron-metrology},
and refs. \cite{Hockett2014,Hockett2015a}), details which were otherwise
lost.

\subsection{Comparison with theoretical results\label{sub:Comparison-with-theory}}

\begin{figure}
\includegraphics{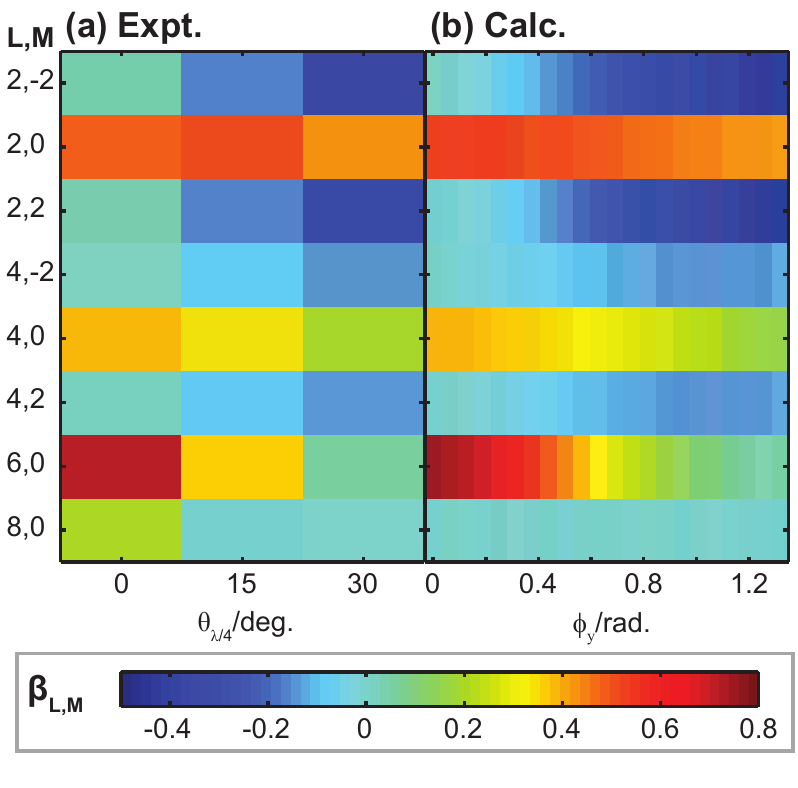}

\caption{Comparison with theoretical results. (a) Experimentally determined
$\beta_{L,M}$, averaged over the main spectral feature, for each
polarization state as defined by $\theta_{\lambda/4}$. (b) Theoretical
results (see papers 1 \& 2 for details), plotted as a function of
polarization state as defined by $\phi_{y}$, the phase shift of the
$y$-polarized component of the electric field. In the former case
$0\leq\theta_{\lambda/4}\leq45^{\circ}$ spans all polarization states
from linear to circular, and in the latter $0\leq\phi_{y}\leq\nicefrac{\pi}{2}$
(the full angular interferogram for $\phi_{y}=0.5$ is shown in figure
\ref{fig:imaging-concepts}). The scales on the plots are set to approximately
match the experimental polarization space. \label{fig:Comparison-with-theory}}

\end{figure}

In ref. \cite{Hockett2014}, a comparison of 3D distributions based
on ionization matrix elements determined from 2D experimental data
were compared qualitatively to tomographically reconstructed experimental
distributions. While qualitatively in good agreement, a more quantitative
comparison at the level of the $\beta_{L,M}(k)$ can now be made.
This comparison is shown in figure \ref{fig:Comparison-with-theory}.
In the theory, the energy-dependence of the $\beta_{L,M}(k)$ is neglected,
hence they are assumed to be constant over the observed radial spectrum.
From the analysis above, it is clear that this assumption is valid
over the FWHM of the main feature in the spectrum, but does not hold
in other regions. Figure \ref{fig:Comparison-with-theory} therefore
compares the theoretical results with the experimental $\beta_{L,M}(k)$
averaged over the FWHM of the main spectral feature. In the theoretical
results the polarization state of the light is parametrized by $\phi_{y}$,
the spectral phase of the $y$-component of the electric field, where
$0\leq\phi_{y}\leq\nicefrac{\pi}{2}$ spans all polarization states
from linear ($\phi_{y}=0$) to circular ($\phi_{y}=\pi/2$). The experimental
results for polarization states $\varepsilon_{1}$, $\varepsilon_{2}$
and $\varepsilon_{3}$ correspond to $\phi_{y}=0$, $\sim$0.5 and
$\sim$1.2~rad. respectively. (For further details of the theory
and a complementary presentation of the theoretical and tomographic
results see ref. \cite{Hockett2015a}.)

It is clear, as expected from the previous qualitative comparison,
that the agreement is good - but not exact. More specifically, the
results show that the tomographic data, as a function of $\theta_{\lambda/4}$,
is close to the theoretical results, plotted as a function of $\phi_{y}$,
over the main spectral peak, and both smoothly evolve with the pulse
polarization. In a general sense this indicates that the theory results,
based solely on 2D data, are validated by this comparison with more
detailed 3D data. It also suggests, however, that some refinement
could be made of the previously obtained ionization matrix elements
based on this more detailed data, in particular via the inclusion
of the radial dependence of the $\beta_{L,M}(k)$ in the fitting methodology.
This is a clear potential offered by the 3D data, and is discussed
further in section \ref{sec:discussion}. 

There are also additional experimental issues which may play a role
here. In this case the exact $\beta_{L,M}(k)$ expansion depends fairly
sensitively on the frame-of-reference definition applied, which includes
both the definition of the image centre and the choice of the $\phi=0$
plane. Appropriate frame-rotations were applied during data analysis,
but small inaccuracies may still be present. Any processing inaccuracies
or artefacts of this type would appear as systematic errors in the
extracted $\beta_{L,M}(k)$, but would not show a clear or smooth
energy dependence, so would not affect the results or conclusions
presented herein more generally. There is also the assumption that
the polarization state of the light is identical over the pulse bandwidth,
which may also not be rigorously true. In this case, an apparent energy
dependence would be observed in the $\beta_{L,M}(k)$, but would be
caused by this polarization state drift rather than any inherent energy
dependence of the photoionization dynamics. Again any effects here
are likely to be small, but may be noticeable at this quantitative
level of comparison. As observed in the previous section, there are
also additional terms which appear in the experimental data, but are
not allowed by the current theoretical treatment. Over the main spectral
feature there is no symmetry-breaking, but significant $L=8$ terms
are present and, as shown in figure \ref{fig:Comparison-with-theory}(b),
do not appear in the theoretical results. The possible origin of these
features is discussed in the following section.

\section*{}

\section{Discussion \& Conclusions\label{sec:discussion}}

In the preceding section the benefits of a full 3D photoelectron measurement
were discussed in terms of the quantitative analysis of the observed
angular interferograms, facilitated by the $\beta_{L,M}(k)$ parameters.
While this treatment is experimentally rigorous, and represents the
full information content of the observable, the complexity of the
coupling of the continuum wavefunction into the observable necessitates
further phenomenological discussion and theoretical analysis in order
to understand the physical significance of these observations, and
draw firmer conclusions about the light-matter interaction.

Firstly, it is important to reiterate that the good agreement of the
current results with the previous model over the main spectral feature
suggest that the dominant channels are identical to the 2D data analysed
in that case (at a lower peak intensity of $\sim$10$^{12}$Wcm$^{-2}$)
and, therefore, the observation of additional interferences in the
3D data are a direct benefit of the maximum information measurement.
It is this enhanced metrology which allows for minor contributions
to the interferogram to be cleanly resolved. The ability to directly
and quantitatively compare the theoretical results from the previous
analysis with experimentally obtained interferograms also provides
a rigorous benchmark against which to validate the previous analysis,
including the accuracy of the ionization matrix elements determined
and the limitations of the theory developed.

In order to understand the additional insights gained from the maximum
information measurements, one can approach from a purely experimental
perspective. In this case, there are two key observations to consider:
the general evolution of the angular interferograms over the energy
spectrum or, equivalently, over the Autler-Townes structure of the
spectrum,%
\footnote{It is of note here that, in the absence of resonance effects, the
ionization matrix elements themselves would not be expected to show
a strong energy dependence over the $\sim$300~meV span of the main
spectral feature considered here. See, for instance, ref. \cite{Park1996}
for discussion.%
} and the strong symmetry-breaking observed in certain energy regions.
These observations are already enough to provide a phenomenological
understanding of the interferences observed. The AT structure is a
direct result of the AC Stark effect, which causes a dynamic shifting
of the energy levels of the ionizing system and continuum as a function
of the laser field. In the frequency-domain, this effect results in
an effective level splitting defined by the time-average of the laser
field, defining a quasi-static ``dressed state'' level structure.
In this picture the photoelectron wavepacket will gain an additional
(energy-dependent) phase due to the level splitting, and interferences
from different components of the AT structure will occur in regions
where photoelectron wavepackets correlated with different quasi-static
levels overlap energetically. Analogous behaviour can be seen in energy-domain
cases where different ionizing transitions are made to interfere via,
for example, multi-colour ionization schemes \cite{Yin1992}, which
similarly allow the creation of new interferences in the continuum.
A similar effect has also been considered in energy-domain work probing
intensity effects, for instance refs. \cite{Dixit1983} and \cite{Ohnesorge1984},
which investigated the high-intensity nano-second pulse regime where
AC Stark shifted levels may be tuned in and out of resonance with
the excitation pulse, yielding a strong intensity dependence in the
angular interferograms according to the number and nature of the states
which were coupled by the multi-photon ionization scheme at a given
intensity. Due to the the additional laser bandwidth present in the
femto or atto-second regime, and consequent broad photoelectron bands,
this type of effect might be expected to be very general, and has
indeed recently been explored in theory for atto-second ionization
in strong fields \cite{Yuan2014}. 

A more appropriate framework in this case is that of a time-domain
interferometer. Here, there is an additional time-dependent phase,
and additional interferences will appear in the measurement - which
is integrated over the pulse duration - providing that the instantaneous
contributions to the continuum wavefunction remain coherent. Examples
of this type of effect include the time-domain interferometer discussed
in ref. \cite{Wollenhaupt2002}, and the ``atomic phase-matching''
in the two-pulse control scheme of ref. \cite{Trallero-Herrero2006}.
In this picture, the additional interferences observed are analogous
to those which appear with polarization-shaped pulses, in which the
coherent temporal interferences are correlated with the instantaneous
polarization state of the pulse, allowing for ``polarization multiplexing''
in the time-integrated measurement (see refs. \cite{Hockett2014,Hockett2015a}).
Here, the additional interferences are associated with the instantaneous
pulse intensity, but are otherwise analogous. These types of interferences
can be generally termed \emph{dynamic}, since they depend on the details
of the laser pulse and the driven dynamics of the ionizing system.
This phenomenology readily explains a strong dependence of the photoelectron
interferograms on the AT structure, and is in fact implicit in the
band structure of the photoelectron energy spectrum which results
from these same dynamical interferences \cite{Wollenhaupt2005,Bayer2008},
and have also recently been explored in the context of intense XUV
pulses \cite{Demekhin2013}. However, this phenomenology does not
obviously account for the observed symmetry breaking which requires
$m$-state dependent phase contributions. This latter effect may,
however, be a result of the polarization dependence of the temporal
phase, either in terms of the bound or continuum states. For instance,
the populations of the $4p_{+1}$ and $4p_{-1}$ bound states accessed
by the left and right circularly polarized components of the field
will correlate with the instantaneous electric field polarization,
hence ionize at different parts of the field cycle and accumulate
different temporal phases.%
\footnote{It is, however, of note that this effect should not be present for
pure linearly polarized pulses, unless some other source of symmetry
breaking is present.%
} In the dressed state (energy-domain) picture this would be manifested
as exactly the required symmetry breaking of the positive and negative
$m$ states, with a time-dependent superposition of the dressed states
created - effectively a manifestation of the electronic ring currents
discussed in the current context in ref. \cite{Wollenhaupt2009a},
in terms of high-harmonic generation in ref. \cite{Xie2008} and more
recently in the atto-second XUV regime in ref. \cite{Yuan2014}.%
\footnote{The spin-orbit components of the $4p$ manifold (with $m=\pm\nicefrac{1}{2},\,\pm\nicefrac{3}{2}$)
may also play a role here \cite{Bayer2008}. Although these were neglected
in previous work (refs. \cite{Hockett2014,Hockett2015a}), since the
ionization dynamics will be identical in both cases aside from the
overall coupling strength into the continuum, the further dynamic
splitting of these states could also allow for the accumulation of
different temporal phases and contribute to the observed $m$-level
symmetry-breaking. Conceptually this is identical to the spin-neglected
case, but results in a more complex 4-state picture. In principle
this is again similar to the energy-domain work of ref. \cite{Dixit1983}.%
} This phase would be directly mapped into the ionization continuum,
again analogously to the polarization multiplexed case previously
explored. Similar considerations may also apply to the continuum states,
in the case where the ionization cannot be considered in the perturbative
regime.

To further investigate the specifics of the light-matter interaction,
and the additional minor contributions to the photoelectron interferograms,
a more quantitative analysis of the observations can be made by considering
the results within our existing theoretical framework. This treatment
assumes a single active electron, dipole coupling and perturbative
ionization (see papers 1 \& 2 for further details). In this scheme
strict limits are placed on the allowed angular momenta and symmetries
of the final continuum states. Any breaking of these limits therefore
indicates additional physical complexities not included in the modelling
of the light-matter coupling, for instance the dynamical phase discussed
above, other non-perturbative effects, multi-electron effects and
so forth. For the 3-photon absorption process outlined in section
\ref{sec:Photoelectron-metrology}, i.e. $4s+h\nu\rightarrow4p_{\pm1}+2h\nu\rightarrow|k,l,m\rangle$,
treated within this scheme, the final states are restricted to $l=1,\,3$
and odd-$m$ terms only. This, in turn, places limits on observable
interferogram: only $\beta_{L,M}(k)$ with even-$L$ and even-$M$
are allowed, and are further restricted to $L\leq6$. Therefore, the
appearance of additional $L,\, M$ terms in the experimental data
indicates the presence of additional partial-waves in the continuum
wavefunction. The dynamic interferences descried above do not involve
angular momentum exchange, so cannot be responsible for the appearance
of new $L$ terms, although they may be implicated in symmetry breaking
if $\pm m$ states are split and accumulate different phases as suggested
above. 

The appearance of additional angular momentum rather indicates additional
interactions - either (a) additional photon absorption and/or (b)
photoelectron (re)scattering. The former could lead to angular momentum
ladder climbing if additional $|n,l,m\rangle$ bound or dressed-states
were accessible at the 1 or 2-photon level (e.g. high-lying Rydbergs),
although it is not clear which specific states could be coupled in
this way, and result in ensemble polarization; a similar effect could
result from multiple, cascaded Raman transitions (ultimately equivalent
to a field-mediated picture), in which atomic orbitals align with
the strong laser field thus creating ensemble alignment. In the case
of elliptically polarized light, such effects could drive electronic
wavepackets with ring-current like behaviour, as suggested above,
and this effect could indeed be responsible for the observed symmetry-breaking
depending on the time-averaged ionization of the ring-current density.
The latter effect essentially describes any other core-photoelectron
interactions, such as angular momentum exchange during ionization,
or field-mediated effects at longer range. In this case the single
active electron picture (essentially a hydrogenic light-matter interaction)
breaks down, and additional electron-electron scattering occurs. Such
processes can be treated with more complex angular momentum coupling
schemes, and experimentally would be indicated by the creation of
electronically excited ions. 

A final, but less likely, possibility is the breakdown of the dipole
approximation. This would also result in the creation of high-order
continuum states, but due to direct multi-polar light-matter couplings.
Since the dipole approximation relies on the wavelength of the electric
field being large compared to the target, it is expected to hold
at 800~nm and low photoelectron energies, although it has previously
been observed that relatively low-energy processes may still require
multipole couplings \cite{Hemmers2003,Lepine2004}. However, such
effects break the interferogram symmetry along the photon propagation
axis, so would not explain the main observations here. Finally it
should be noted that additional macroscopic effects may also play
a role here, including intensity averaging over the laser pulse and
the static electric fields present in the VMI chamber (which are known
to influence states near threshold as mentioned in sect. \ref{sub:3D-photoelectron-momentum-images});
any irregularities in the laser pulse could also affect the minor
channels observed, such as non-uniform spectral phase and polarization
state, although such effects are expected to be negligible in this
case. 

From these considerations it appears that both dynamical interferences,
resulting from the mapping of the AT effect onto the photoelectron
spectral phase and resulting in a strong energy dependence to the
interferograms, and angular momentum couplings, resulting in the $L>6$
terms observed in the interferograms, must both be present. Although
a more sophisticated theoretical treatment, and possibly additional
experiments to consider the intensity dependence, is needed in order
to asses the exact nature of these possibilities, and the precise
details of the symmetry-breaking interactions, from the experimental
perspective the ability to \emph{resolve} these effects is a uniquely
powerful result of maximum information methods. The 3D data, reflecting
the complex set of interfering ionization channels through the observable
dissected in terms of angular interferograms and associated anisotropy
parameters, reveals a wealth of detailed information on the fundamental
physics of the light-matter interaction, and scattering of the outgoing
electron, and requires no \emph{a priori} assumptions regarding the
symmetry of the light-matter interaction. This high level of detail
necessitates careful analysis but, ultimately, provides the most complete
picture of such interactions possible, whether the goal is quantum
metrology or quantum control.

Financial support by the State Initiative for the Development of Scientific
and Economic Excellence (LOEWE) in the LOEWE-Focus ELCH is gratefully
acknowledged.

\bibliographystyle{unsrt}
\bibliography{/media/hockettp/StoreM/reports/bibliography/baumert_paper3_151014_problemRefs,/media/hockettp/StoreM/reports/bibliography/baumert_paper3_040315}

\end{document}